\documentclass[%
 reprint,onecolumn,
 amsmath,amssymb,
 aps,
]{revtex4-2}

\usepackage{graphicx}
\usepackage{dcolumn}
\usepackage{bm}
\usepackage{epstopdf}
\usepackage{braket}
\usepackage{tcolorbox}

\begin{document}

\preprint{APS/123-QED}

\title{Quantum optics in MATLAB}
\author{Nilakantha Meher}
\email{nilakantha.meher6@gmail.com}
\affiliation{Department of Chemical and Biological Physics,
Weizmann Institute of Science, Rehovot 7610001, Israel}

\begin{abstract}
We provide a MATLAB numerical guide at the beginner level to support students starting their research careers in theoretical quantum optics and related areas. These resources are also valuable for undergraduate and graduate students working on semester projects in similar fields.
\end{abstract}

%

%

%
\maketitle
\tableofcontents
%

\section{Introduction}
MATLAB is a user-friendly and robust framework for numerical computing based on matrix operations. Several numerical toolboxes or open-source packages written in MATLAB \cite{Tan1999JOP,korsch2016computations,schmidt2017wavepacket,schmidt2018wavepacket,
chelikowsky2019introductory, norambuena2020coding} have been designed to address
analytically intractable problems in quantum mechanics, quantum optics and condensed matter physics.

In this tutorial, we present various numerical codes written in MATLAB to help students understand the basics of quantum optics. These codes can be easily extended to address a wide range of research problems in quantum optics and related areas that involve high-dimensional matrix manipulations. Importantlly, they can be executed in any version of MATLAB without requiring pre-installed packages.\\ 

\noindent
Before we start writing codes in MATLAB, we follow the rules:
\begin{itemize}
\item Pure quantum states are represented by normalized column matrices with unit norm, while mixed states are represented by square matrices with unit trace. 
\item Quantum operators are represented by square matrices.
\item We need to pre-decide the dimension of the matrices for states and operators based on the specific problem at hand. The dimension refers to the number of rows or columns. If your results deviate from your expectations, you can consider increasing the dimension of the matrices. We will learn how to decide the appropriate dimensions of the matrices as we write code. 
\item  We use \textquoteleft clear\textquoteright~ to clear memory and \textquoteleft clc\textquoteright~ to clear the output screen at the beginning of a code to prevent any numerical errors.
\item To suppress output, add a semicolon \textquoteleft ;\textquoteright~ at the end of the respective line.
\item The symbol \% is used to comment a line. The sentence after \% is used to explain the code to the readers. It will not be executed in the run of a program and will not be displayed at the output.
\item We use \textquotedblleft format short" for truncating the number of digits after the decimal point to be four. One may choose \textquotedblleft formal long" to display more number of digits after the decimal point.
\item We listed a few commands in Table. \ref{Table} that we use to perform calculations.  
\end{itemize}
\begin{table}\label{Table}
\begin{center}
\begin{tabular}{|c|c|}
\hline
Operations & MATLAB commands\\
\hline
$a\times b$   & $a*b$\\
\hline
$\sqrt{a}$ & sqrt($a$)\\
\hline
$e^a$ & exp($a$)\\
\hline
$a^b$ & $a\hat~b$\\
\hline
$n!$ & prod(1:$n$)\\
\hline
Norm of a column vector/matrix X  & norm(X)\\
\hline
Identity matrix of dimension $d$ & eye(d)\\
\hline
$e^A$ (exponential of matrix A) & expm(A)\\
\hline
Trace of a matrix A & trace(A)\\
\hline
Eigenvectors and eigenvalues  of A & [Evec, Eval]=eig(A)\\
\hline
$n$th column of a matrix A & A(:,n)\\
\hline
A$\otimes$B (tensor product) & kron(A,B)\\
\hline
\end{tabular}
\caption{Used operations and their MATLAB commands. }
\end{center}
\end{table}
\section{Quantum states in MATLAB}
\subsection{Number States}
The number states are the most fundamental quantum states of a quantized electromagnetic (EM) field. The number state $\ket{n}$ represents a field having $n$ photons. 

In MATLAB, we represent a number state $\ket{n}$ with a column matrix in which the element at the ($n+1$)th position is 1, while the other elements are 0. Accordingly, the vacuum state $\ket{0}$ is represented by a column matrix in which the first element is 1 and the others are 0.\\

\noindent
\textbf{Trick:} Consider each column of an identity matrix to represent a number state.\\

\noindent
\textbf{{Code for number states (vacuum state, single-photon state, and two-photon state):}}
\begin{verbatim}
clear;  	   % Clear memory
clc;    	   % Clear the command window/screen
d=5;  		   %dimension of the field 
I=eye(d); 	   %Identity matrix of dimension d
Vacuum=I(:,1)  %First column of identity matrix: vacuum state
Ket1=I(:,2)    %Second column of identity matrix: single-photon state |1⟩
Ket2=I(:,3)    %Third column of identity matrix: two-photon state |2⟩
\end{verbatim}
The outputs are:
\begin{verbatim}
Vacuum=         Ket1=         Ket2=
        1              0            0
        0              1            0
        0              0            1
        0              0            0
        0              0            0
\end{verbatim}
It is to be noted that the above column matrices are derived from a 5$\times$5 identity matrix (d=5). To create higher-order number states, we need to use a larger value for \textquoteleft d\textquoteright. For example, if we wish to create a photon-number state 20, \textquoteleft d\textquoteright~ must be chosen greater than 20.\\ 

\noindent
\textbf{Code for number state with photon number 20:}
\begin{verbatim}
clear;  % Clear memory
clc;    % Clear the command window/screen
d=21;  %dimension of the field 
I=eye(d); % Identity matrix of dimension d
Ket20=I(:,21) % number state |20⟩
\end{verbatim}
The output is a column matrix with 21st element is 1 and others are zero. The configuration represents the number state $\ket{20}$.


\subsection{Superposition of Number States}
In the previous subsection, we mapped the number states to columns of an identity matrix. For example, $n$-photon state $\ket{n}$ is mapped to $(n+1)$th column of the identity matrix. Now, using this mapping, let us we write the following superposition state in MATLAB:
\begin{align}\label{Superposed}
\ket{\psi}=\frac{1}{\sqrt{3}}\ket{2}+\frac{1}{\sqrt{2}}\ket{5}-\frac{1}{\sqrt{6}}\ket{6}.
\end{align}
We must ensure that the state $\ket{\psi}$ is normalized. The norm of $\ket{\psi}$ is calculated to be $|1/\sqrt{3}|^2+|1/\sqrt{2}|^2+|-1/\sqrt{6}|^2=1$, and hence, $\ket{\psi}$ is normalized.\\ 
\noindent
\textbf{Code for creating a superposed state:}
\begin{verbatim}
clear;  % Clear memory
clc;    % Clear the command window/screen
d=7;  		%dimension of the field 
I=eye(d); 	% Identity matrix of dimension d
Ket2=I(:,3); % two-photon state |2⟩
Ket5=I(:,6); % five-photon state |5⟩
Ket6=I(:,7); % six-photon state |6⟩
Psi=1/sqrt(3)*Ket2+1/sqrt(2)*Ket5-1/sqrt(6)*Ket6      %superposition state
N_psi=norm(Psi)              %norm of the state Psi
\end{verbatim}
The outputs are:
\begin{verbatim}
Psi =
         0
         0
    0.5774
         0
         0
    0.7071
   -0.4082
   
N_psi =
     1
\end{verbatim}
As N\_psi=1, the state is normalized. The output state is a column matrix in which the 3rd, 6th and 7th elements are non-zero, which correspond to the amplitudes of the number states $\ket{2}$, $\ket{5}$, and $\ket{6}$ in the superposition state, respectively. As the largest number-state present in the superposition state is $|6\rangle$, the dimension (d) of the field is taken to be 7. In general, if a superposition state contains the largest number-state to be $|n\rangle$, then \textquoteleft d' must be larger than $n+1$.  \\

\noindent
\textit{\textbf{Note: Any superposition of number states can be expressed as a linear combination of the columns of an identity matrix. The dimension (d) of the matrix must be larger than the largest number-state present in the superposition state. }}
\subsection{Coherent States}
A coherent state $\ket{\alpha}$ is a superposition of all number states \cite{Glauber1963PR, Klauder1963JMP}. In the number basis, it is expressed as [Ch. 3 of Ref. \cite{Gerry}]
\begin{align}
\ket{\alpha}=e^{-|\alpha|^2/2}\sum_{n=0}^\infty \frac{\alpha^n}{\sqrt{n!}} \ket{n}.
\end{align}
Therefore, to write this state in MATLAB, we need to sum all the columns of an identity matrix weighted by the coefficient $e^{-|\alpha|^2/2}\frac{\alpha^n}{\sqrt{n!}}$. In principle, one needs to add infinite number of such column matrices. However, for a given $\alpha$, the coefficient $e^{-|\alpha|^2/2}\frac{\alpha^n}{\sqrt{n!}}$ becomes negligible as $n$ increases. Thus, we truncate the sum so the norm of the state remains very close to 1. \\ 

\noindent
\textbf{Code for coherent state:}
\begin{verbatim}
clear;  % Clear memory
clc;    % Clear the command window/screen
d=10; %dimension of the field
I=eye(d); 
alpha=0.6; %Amplitude of the coherent state
Coh=0;  %initialization
for x=0:d-1;
Coh=Coh+exp(-norm(alpha)^2/2)*alpha^x/sqrt(prod(1:x))*I(:,x+1);
end
Coh    %it will display the coherent state
N_c=norm(Coh)  %checking norm 
\end{verbatim} 
The outputs are:
\begin{verbatim}
Coh=
    0.8353
    0.5012
    0.2126
    0.0737
    0.0221
    0.0059
    0.0015
    0.0003
    0.0001
    0.0000
    
N_c=
    1.0000   
\end{verbatim}
Although we have truncated the above state up to d=10 (or $n=9)$, it is important to note that the norm N\_c of the coherent state is very close to 1. As we use \textquotedblleft format short", the output displays N\_c=1.0000. If we use \textquotedblleft format long", then it will display N\_c=0.999999999996367, which is very close to 1.  

If we take a larger value of \textquoteleft alpha\textquoteright, the state with d=10 may not have the unit norm. For that case, we need to increase the value of \textquoteleft d\textquoteright. \\   

\noindent
\textit{\textbf{Note:} If the norm of a state is found to be less than 1, it is necessary to increase the dimension of the field \textquoteleft d\textquoteright. Please refer to \ref{Dimension} for an explanation of the significance of dimension \textquoteleft d\textquoteright.  It illustrates the amount of error that may arise when choosing a smaller value for \textquoteleft d\textquoteright. }

\subsection{Thermal States}
The electromagnetic radiation from an object at a non-zero temperature is a thermal light. The density matrix of a thermal state in the number basis is [Ch. 2 of Ref. \cite{Gerry}] 
\begin{align}
\rho_{th}=\frac{1}{1+n_{th}}\sum_{n=0}^\infty \left(\frac{n_{th}}{1+n_{th}} \right)^n \ket{n}\bra{n},
\end{align}
where $n_{th}$ is the average number of photons in the thermal state $\rho_{th}$.

As thermal states are mixed, we represent them by a square matrices, and they are diagonal in number basis. As we can see from the above equation, it is a sum of square matrices $\ket{n}\bra{n}$ for all $n$ with appropriate coefficients (probabilities). To get the matrix form of $\ket{n}\bra{n}$, consider an example. For $n=3$, we have
\begin{equation}
\ket{3}\bra{3}=\left(\begin{array}{c}
0\\
0\\
0\\
1\\
0\\
\vdots
\end{array}\right)\left(\begin{array}{cccccc}
0 & 0 & 0 & 1 & 0 & \cdots
\end{array}\right)=\left( \begin{array}{cccccc}
0 &0 & 0 & 0 & 0 &\cdots\\
0 & 0 & 0 & 0 & 0 &\cdots\\
0 & 0 & 0 & 0 & 0 & \cdots\\
0 & 0 & 0 & 1 & 0 & \cdots\\
0 & 0 & 0 & 0 & 0 &\cdots\\
\vdots & \vdots & \vdots & \vdots &\vdots &\ddots\\
\end{array} \right).
\end{equation}
It is a square matrix whose 4th element of the diagonal is 1.\\

\noindent
\textbf{Code for thermal state:}
\begin{verbatim}
clear;  % Clear memory
clc;    % Clear the command window/screen
d=10;
I=eye(d);
nth=0.5; %Average number of photons in thermal state
RhoTh=0; %Initialization
for x=0:d-1
RhoTh=RhoTh+nth^(x)/(1+nth)^(x+1)*I(:,x+1)*I(:,x+1)';
end
RhoTh  %it will display thermal state
N_th=trace(RhoTh)   %checking trace to be 1
\end{verbatim}
The outputs are:

\begin{verbatim}
RhoTh=
0.6667       0       0       0       0       0       0       0      0       0
     0  0.2222       0       0       0       0       0       0      0       0
     0       0  0.0741       0       0       0       0       0      0       0
     0       0       0  0.0247       0       0       0       0      0       0
     0       0       0       0  0.0082       0       0       0      0       0
     0       0       0       0       0  0.0027       0       0      0       0
     0       0       0       0       0       0  0.0009       0      0       0
     0       0       0       0       0       0       0  0.0003      0       0
     0       0       0       0       0       0       0       0 0.0001       0
     0       0       0       0       0       0       0       0      0  0.0000

N_th =
    1.0000
\end{verbatim}

\subsection{Squeezed Vacuum States}
A squeezed vacuum state in the number basis is [Ch. 7 of Ref. \cite{Gerry}]
\begin{align}
\ket{\xi}=\frac{1}{\sqrt{\cosh r}} \sum_{n=0}^\infty (-1)^n \frac{\sqrt{(2n)!}}{2^n n!}e^{in\theta} (\tanh r)^n \ket{2n},
\end{align}
where $r$ and $\theta$ are the squeeze parameters. This state is a superposition of all the even number states.\\

\noindent
\textbf{Code for squeezed vacuum state:}
\begin{verbatim}
clear;  % Clear memory
clc;    % Clear the command window/screen
d=20;
I=eye(d);
r=0.3; 				%squeezing parameter
theta=pi/4;         %squeezing direction
Sqz=0;              %initialization
for x=0:(d/2)-1;
p=(1/sqrt(cosh(r)))*sqrt(prod(1:(2*x)))/(2^x*prod(1:x));
Sqz=Sqz+p*(-1)^x*exp(i*x*theta)*(tanh(r))^x*I(:,2*x+1);
end
Sqz               	%squeezed vacuum state output
N_sqz=norm(Sqz)   	%checking norm to be 1
\end{verbatim}
The outputs are:
\begin{verbatim}
Sqz =
   0.9781 + 0.0000i
   0.0000 + 0.0000i
  -0.1425 - 0.1425i
   0.0000 + 0.0000i
   0.0000 + 0.0508i
   0.0000 + 0.0000i
   0.0096 - 0.0096i
   0.0000 + 0.0000i
  -0.0037 + 0.0000i
   0.0000 + 0.0000i
   0.0007 + 0.0007i
   0.0000 + 0.0000i
  -0.0000 - 0.0003i
   0.0000 + 0.0000i
  -0.0001 + 0.0001i
   0.0000 + 0.0000i
   0.0000 - 0.0000i
   0.0000 + 0.0000i
  -0.0000 - 0.0000i
   0.0000 + 0.0000i
   
N_sqz =
    1.0000  
\end{verbatim}
\subsection{Number State Filtered Coherent State}
A number state filtered coherent state (NSFCS) in number basis is \cite{Meher2018QInP}
\begin{align}
\ket{\psi(\alpha,m)}=\frac{e^{-|\alpha|^2/2}}{\sqrt{N_m}}\sum_{n=0,n\neq m}^\infty \frac{\alpha^n}{\sqrt{n!}} \ket{n},
\end{align}
where the normalization constant $N_m=1-e^{-|\alpha|^2}\tfrac{|\alpha|^{2m}}{{m!}}$.
This definition implies that the state is obtained if the number state $\ket{m}$ is absent in the superposition. \\

\noindent
\textbf{Code for NSFCS:}
\begin{verbatim}
clear;  % Clear memory
clc;    % Clear the command window/screen
d=15;
m=4; %number state 4 will be absent in the distribution
I=eye(d);
alpha=0.8; %Amplitude of the coherent state
nsfs=0;    %initialization
for x=0:d-1;
if x==m
nsfs=nsfs+0*I(:,x+1);
else
nsfs=nsfs+exp(-norm(alpha)^2/2)*alpha^x/sqrt(prod(1:x))*I(:,x+1);
end
end
NSFS=nsfs/norm(nsfs)   %this is to normalize the state            
\end{verbatim}
The output is:

\begin{verbatim}
NSFS =
    0.7275
    0.5820
    0.3292
    0.1521
         0
    0.0218
    0.0071
    0.0021
    0.0006
    0.0002
    0.0000
    0.0000
    0.0000
    0.0000
    0.0000
\end{verbatim}
Note that the 5th element is zero indicating the absent of the number state $\ket{4}$. 
\subsection{States of a two-level atom}
Consider a two-level atom whose excited and ground states are represented by $\ket{e}$ and $\ket{g}$ respectively. We represent these states by column matrices of dimension 2. We have
\begin{align}
\ket{e}=\left(\begin{array}{c}
1\\
0
\end{array}\right), \ket{g}=\left(\begin{array}{c}
0\\
1
\end{array}\right).
\end{align} 
Then, any superposition of these states is
\begin{align}
\ket{\psi}=\alpha \ket{e}+\beta\ket{g},
\end{align}
such that $|\alpha|^2+|\beta|^2=1$ (normalization condition).\\

\noindent
\textbf{Code for atomic states:}
\begin{verbatim}
clear;  % Clear memory
clc;    % Clear the command window/screen
es=[1;0]  %excited state
gs=[0;1]  %ground state
alpha=sqrt(0.4); %superposition coefficient
beta=sqrt(0.6); %superposition coefficient
Psi=alpha*es+beta*gs   %superposition state
\end{verbatim}
The outputs are:
\begin{verbatim}
es =           gs=             Psi=
     1           0                 0.6325
     0           1                 0.7746
\end{verbatim}
We check that the norm of \textquoteleft Psi' is $0.6325^2+0.7746^2= 1$. 
\section{Operators in MATLAB}
The operators correspond to a quantized EM field will be written in number basis ($\ket{n}$) and the operators correspond to two-level atoms will be written in atomic state basis ($\ket{e}$ and $\ket{g}$).

\subsection{Annihilation, Creation and Number operators} 
The basic operators that we use to study the EM field in quantum optics are the annihilation operator, creation operator and the number operator \cite{Scully}. To represent these operators in matrix form, we first need to know their action on number states. 

The action of annihilation and creation operators on a number state is
\begin{subequations}
\begin{align}
\hat a\ket{n}&=\sqrt{n}\ket{n-1},\\
\hat a^\dagger \ket{n}&=\sqrt{n+1}\ket{n+1}.
\end{align}
\end{subequations}
The action of annihilation operator on a field state corresponds to the destruction of a photon from the field, while the action of creation operator corresponds to the creation of a photon in the field. The above two equations give the matrix form
\begin{align}
\hat a=\left( \begin{array}{ccccc}
0 &\sqrt{1} & 0 & 0 &\cdots\\
0 & 0 & \sqrt{2} & 0 &\cdots\\
0 & 0 & 0 & \sqrt{3} &\cdots\\
0 & 0 & 0 & \ddots &\cdots\\
\vdots & \vdots & \vdots & \vdots &\ddots\\
\end{array} \right),\hat a^\dagger=\left( \begin{array}{ccccc}
0 & 0 & 0 & 0 &\cdots\\
\sqrt{1} & 0 & 0 & 0 &\cdots\\
0 & \sqrt{2} & 0 & 0 &\cdots\\
0 & 0 & \sqrt{3} & \ddots &\cdots\\
\vdots & \vdots & \vdots & \vdots &\ddots\\
\end{array} \right).
\end{align}
\textit{\textbf{Note:} The creation operator is adjoint of the annihilation operator.}\\

The number operator is $\hat N=\hat a^\dagger \hat a$, and its action on number state is
\begin{align}
\hat N\ket{n}=n\ket{n}.
\end{align}
Thus, it is diagonal in number basis. The expectation value of the number operator in a number state counts the number of photons in that state, that is,
\begin{align}
\bra{n}\hat N\ket{n}=n.
\end{align} 

\noindent
\textbf{Code for annihilation, creation and number operators:}
\begin{verbatim}
clear;  % Clear memory
clc;    % Clear the command window/screen
d=6;      %dimension of the operator
A = diag(sqrt(1:d-1), 1)  % Annihilation operator
Ad=A'         %Creation operator (adjoint of annihilation operator)
N=Ad*A        %Number operator
\end{verbatim}
The outputs are:
\begin{verbatim}
A =
         0    1.0000         0         0         0         0
         0         0    1.4142         0         0         0
         0         0         0    1.7321         0         0
         0         0         0         0    2.0000         0
         0         0         0         0         0    2.2361
         0         0         0         0         0         0
Ad =
         0         0         0         0         0         0
    1.0000         0         0         0         0         0
         0    1.4142         0         0         0         0
         0         0    1.7321         0         0         0
         0         0         0    2.0000         0         0
         0         0         0         0    2.2361         0
N =
         0         0         0         0         0         0
         0    1.0000         0         0         0         0
         0         0    2.0000         0         0         0
         0         0         0    3.0000         0         0
         0         0         0         0    4.0000         0
         0         0         0         0         0    5.0000
\end{verbatim}

\subsection{Hamiltonian for the electromagnetic field}
The Hamiltonian for a quantized electromagnetic field is [Ch. 2 of Ref. \cite{Gerry}, Ch. 1 of Ref. \cite{Scully}]
\begin{align}
\hat H=\left(\hat a^\dagger \hat a+\frac{1}{2}\hat I_f\right)\hbar\omega,
\end{align}
where $\hbar=6.626\times 10^{-34}$Js (Planck's constant) and $\omega$ is the frequency of the field. The operator $\hat I_f$ is an identity operator. In many textbooks, authors don't explicitly write the identity operator $\hat I_f$. \\ 

\noindent
\textit{\textbf{Note:} We often take $\hbar\omega=1$ to normalize the output quantities in the unit of energy. }\\

\noindent  
\textbf{Code for EM field Hamiltonian:}
\begin{verbatim}
clear;  	% Clear memory
clc;    	% Clear the command window/screen
d=6;       	%dimension of the operator
I=eye(d);   	%identity matrix    
hbar=1;
omega=1;    	%frequency of the cavity field
A = diag(sqrt(1:d-1), 1);  	% Annihilation operator
Ad=A';		%Creation operator (adjoint of annihilation operator)
H=hbar*omega*(Ad*A+(1/2)*I)  	%Hamiltonian        
\end{verbatim}
The output is:
\begin{verbatim}
H =
    0.5000         0         0         0         0         0
         0    1.5000         0         0         0         0
         0         0    2.5000         0         0         0
         0         0         0    3.5000         0         0
         0         0         0         0    4.5000         0
         0         0         0         0         0    5.5000
\end{verbatim}
 
\subsection{Pauli matrices for two-level atom} 
Let the excited and ground states of a two-level atom, represented by $\ket{e}$ and $\ket{g}$, have the energies $\hbar\omega_0/2$ and $-\hbar\omega_0/2$ respectively. Thus, the energy difference is $\hbar\omega_0$. The Hamiltonian for a two-level atom is  \cite{raimond2001manipulating}
\begin{align}
\hat H=\frac{\hbar\omega_0}{2}\hat \sigma_z,
\end{align} 
where 
\begin{align}
\hat\sigma_z=\left(\begin{array}{cc}
1 & 0\\
0 & -1
\end{array}\right), 
\end{align}
is a Pauli matrix. To account for the transition between the two energy levels, we consider the raising and lowering operators
\begin{align}
\hat\sigma_+=\left(\begin{array}{cc}
0 & 1\\
0 & 0
\end{array}\right),~
\hat\sigma_-=\left(\begin{array}{cc}
0 & 0\\
1 & 0
\end{array}\right), 
\end{align}
respectively. Their actions are
\begin{align}
\hat\sigma_+\ket{g}=\ket{e},
\hat\sigma_-\ket{e}=\ket{g}.
\end{align}
 
\noindent
\textit{\textbf{Note:} We often take $\hbar\omega_0=1$ to normalize the output quantities in the unit of energy. }\\

\noindent
\textbf{Code for atomic Hamiltonian, raising and lowering operators:}
\begin{verbatim}
clear;  % Clear memory
clc;    % Clear the command window/screen
hbar=1;
Omega_0=1;     %Atomic transition frequency
Sz=[1 0;0 -1]								%Sigma_z operator
Splus=[0 1;0 0]							%Raising operator
Sminus=[0 0;1 0]							%Lowering operator
H=hbar*Omega_0/2*Sz								%Hamiltonian for a two-level atom
\end{verbatim}
The outputs are
\begin{verbatim}
Sz =
     1     0
     0    -1
Splus =
     0     1
     0     0
Sminus =
     0     0
     1     0
H =
    0.5000         0
         0   -0.5000
\end{verbatim}

\section{Properties of quantum states}
\subsection{Photon number distribution}\label{ProbabilityDistribution}
Upon measurement on a number state $\ket{n}$, we obtain $n$ photons with unit probability. However, there are several nonclassical fields that do not possess a definite number of photons \cite{klyshko1996nonclassical}. When we measure the photon number in such fields, we collapse the field state to a particular number state with a definite number of photons. If we perform repeated measurements on an ensemble of such field states, we encounter different photon numbers with associated probabilities. For example, the photon-number probability distribution in a coherent state is Poissonian. 
   
The probability of finding $n$ photons in a given field state $\ket{\psi}$ is 
\begin{align}
P_n=|\langle n|\psi\rangle|^2.
\end{align}
In particular, the probability of detecting $n$ photons in a coherent state $\ket{\alpha}$ is
\begin{align}
P_n=|\langle n|\alpha\rangle|^2=e^{-|\alpha|^2}\frac{|\alpha|^{2n}}{{n!}},
\end{align}
and in a thermal state is [Ch. 2 of Ref. \cite{Gerry}]
\begin{align}
P_n=\langle n|\rho_{th}|n\rangle=\frac{n_{th}^n}{(1+n_{th})^{(1+n)}} .
\end{align}

\begin{figure}
\includegraphics[scale=0.5]{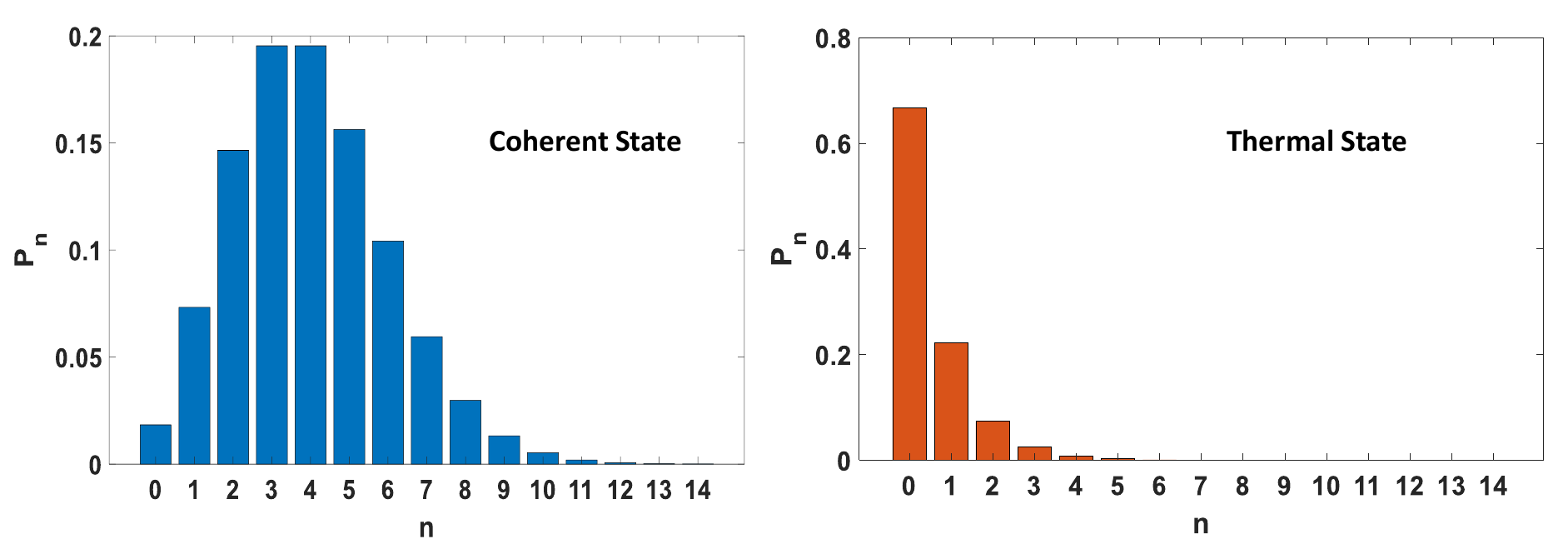}
\caption{Photon number distribution $P_n$ for coherent state $\ket{\alpha}$ with $\alpha=2$ (left) and for thermal state $\rho_{th}$ with $n_{th}=0.5$ (right).}\label{PhotonDist}
\end{figure}

\noindent
\textbf{Code for photon-number probability distributions in coherent and thermal states:}

\begin{verbatim}
clear;  % Clear memory
clc;    % Clear the command window/screen
d=15; %dimension of the field
I=eye(d);
%%%%%%%% Coherent State
alpha=2; %Amplitude of the coherent state
Coh=0;  %initialization
for x=0:d-1;
Coh=Coh+exp(-norm(alpha)^2/2)*alpha^x/sqrt(prod(1:x))*I(:,x+1);
end

for n=0:d-1
  PnC(n+1)=norm(I(:,n+1)'*Coh)^2;   %Probability of |n> in coherent state
end

%%%%%%%% Thermal state
nth=0.5; %Average number of photons in thermal state
RhoTh=0;
for x=0:d-1
RhoTh=RhoTh+nth^(x)/(1+nth)^(x+1)*I(:,x+1)*I(:,x+1)';
end

for n=0:d-1
  PnTh(n+1)=I(:,n+1)'*RhoTh*I(:,n+1);   %Probability of |n> in thermal state
end

%%%%%%%%%% Bar plots
n=0:d-1;
figure(1)
bar(n,PnC)

figure(2)
bar(n,PnTh)
\end{verbatim} 
The output bar plots are given in Fig. \ref{PhotonDist}. The height of a bar indicates the probability of detecting a corresponding number state in coherent state (left) and thermal state (right).

\subsection{Average number of photons}
The average number of photons in a given field state $\ket{\psi}$ is 
\begin{align}
\langle a^\dagger a\rangle=\bra{\psi}a^\dagger a \ket{\psi}.
\end{align}
For number state $\ket{n}$,
\begin{align}
\langle a^\dagger a\rangle=\bra{n}a^\dagger a \ket{n}=n,
\end{align}
which is the number of photons in the state.

The average number of photons in a coherent state is  \cite{Gerry}
\begin{align}
\langle a^\dagger a\rangle=\bra{\alpha}a^\dagger a \ket{\alpha}=|\alpha|^2,
\end{align}
and in a thermal state is 
\begin{align}
\langle a^\dagger a\rangle=\text{Tr}(a^\dagger a \rho_{th})=n_{th}.
\end{align}

\noindent
\textbf{Code for calculating the average number of photons:}

\begin{verbatim}
clear;  % Clear memory
clc;    % Clear the command window/screen
d=20;  		%dimension of the field 
I=eye(d);   %identity matrix
%%%% create number operator
A = diag(sqrt(1:d-1), 1);                	% Annihilation operator
Ad=A';         								%Creation operator
AdA=Ad*A;                                     %Number operator

%% Create a number state
Ket4=I(:,5);   % four-photon state |4⟩
AdANumber=Ket4'*AdA*Ket4  %Average number of photons in the number state

%% Create a coherent state 
alpha=sqrt(3); %Amplitude of the coherent state
Coh=0;  %initialization
for x=0:d-1;
Coh=Coh+exp(-norm(alpha)^2/2)*alpha^x/sqrt(prod(1:x))*I(:,x+1);
end
AdACoherent=Coh'*AdA*Coh    %Average number of photons in coherent state

%%% Create a thermal state
nth=0.85; %Assumed average number of photons in thermal state
RhoTh=0;
for x=0:d-1
RhoTh=RhoTh+nth^(x)/(1+nth)^(x+1)*I(:,x+1)*I(:,x+1)';
end
AdAThermal=trace(AdA*RhoTh)    % Average number of photons in thermal state
\end{verbatim}

The outputs are:
\begin{verbatim}
AdANumber =
     4
AdACoherent =
    3.0000
AdAThermal =
    0.8500
\end{verbatim}

In the above code, we selected the number state $\ket{4}$ and calculated the average number of photons to be 4. This outcome illustrates that the number operator counts the number of photons in a number state. This is because the number states are eigenstates of the number operator. For coherent state, we take the amplitude to be $\alpha=\sqrt{3}$ and calculated the average number of photons to be $|\alpha|^2=3$. Similarly, we assumed $n_{th}$ to be 0.85 to create thermal state in the code and that is reflected at the output.

\subsection{Zero time-delay second-order coherence function}
The zero time-delay second-order coherence function characterizes the photon statistics of a field \cite{Glauber1963PR,RevModPhys.37.231}. It is defined as [Ch. 5 of Ref. \cite{Gerry}]
\begin{align}
g^{(2)}(0)=\frac{\langle a^{\dagger 2} a^2 \rangle}{\langle a^\dagger a\rangle^2},
\end{align} 
where $\langle a^{\dagger 2} a^2 \rangle=\bra{\psi}a^{\dagger 2} a^2 \ket{\psi}$ and $\langle a^{\dagger } a \rangle=\bra{\psi}a^{\dagger } a \ket{\psi}$.\\ 
\noindent
The $g^{(2)}(0)$ for coherent state $\ket{\alpha}$ is 
\begin{align}
g^{(2)}(0)=1,
\end{align}
indicating Poissonian photon statistics. 

A field with $g^{(2)}(0)<1$ exhibits sub-Poissonian photon statistics, and a field with $g^{(2)}(0)>1$ possesses super-Poissonian photon statistics. For number state $\ket{n}$,
\begin{align}
g^{(2)}(0)=1-(1/n),~~~ \text{(sub-Poissonian)} 
\end{align}
and for thermal state $\rho_{th}$, 
\begin{align}
g^{(2)}(0)=2. ~~~~ \text{(super-Poissonian)}
\end{align}

\noindent
\textbf{Code for calculating $g^{(2)}(0)$:}

\begin{verbatim}
clear;  % Clear memory
clc;    % Clear the command window/screen
d=25;  		%dimension of the field 
I=eye(d);   %identity matrix
A = diag(sqrt(1:d-1), 1);                	% Annihilation operator
Ad=A';         								%Creation operator
AdA=Ad*A;                                     %Number operator
Ad2A2=Ad*Ad*A*A;                            %A-dagger-square A-square

%% Create a number state
Ket4=I(:,5);   % four-photon state |4⟩
AdANumber=Ket4'*AdA*Ket4;     
Ad2A2Number=Ket4'*Ad2A2*Ket4;
g2Number=Ad2A2Number/(AdANumber)^2          % g2(0) for number state

%% Create a coherent state 
alpha=sqrt(3); %Amplitude of the coherent state
Coh=0;  %initialization
for x=0:d-1;
Coh=Coh+exp(-norm(alpha)^2/2)*alpha^x/sqrt(prod(1:x))*I(:,x+1);
end

AdACoherent=Coh'*AdA*Coh;  
Ad2A2Coherent=Coh'*Ad2A2*Coh;  
g2Coherent=Ad2A2Coherent/(AdACoherent)^2     % g2(0) for coherent state

%%% Create a thermal state
nth=0.85; %Average number of photons in thermal state
RhoTh=0;
for x=0:d-1
RhoTh=RhoTh+nth^(x)/(1+nth)^(x+1)*I(:,x+1)*I(:,x+1)';
end

AdAThermal=trace(AdA*RhoTh); 
Ad2A2Thermal=trace(Ad2A2*RhoTh);   
g2Thermal=Ad2A2Thermal/(AdAThermal)^2      %g2(0) for thermal state
\end{verbatim}
\vspace{1cm}
The outputs are:
\begin{verbatim}
g2Number =
    0.7500
g2Coherent =
    1.0000
g2Thermal =
    2.0000
\end{verbatim} 

Note that, we take the number state to be $\ket{4}$ and we got $g^{(2)}(0)=1-(1/4)=0.75$. The $g^{(2)}(0)$ for coherent and thermal states are 1 and 2, respectively, independent of the average photon number. 
\section{Atom-Field Interaction}
\begin{figure}
\includegraphics[scale=0.5]{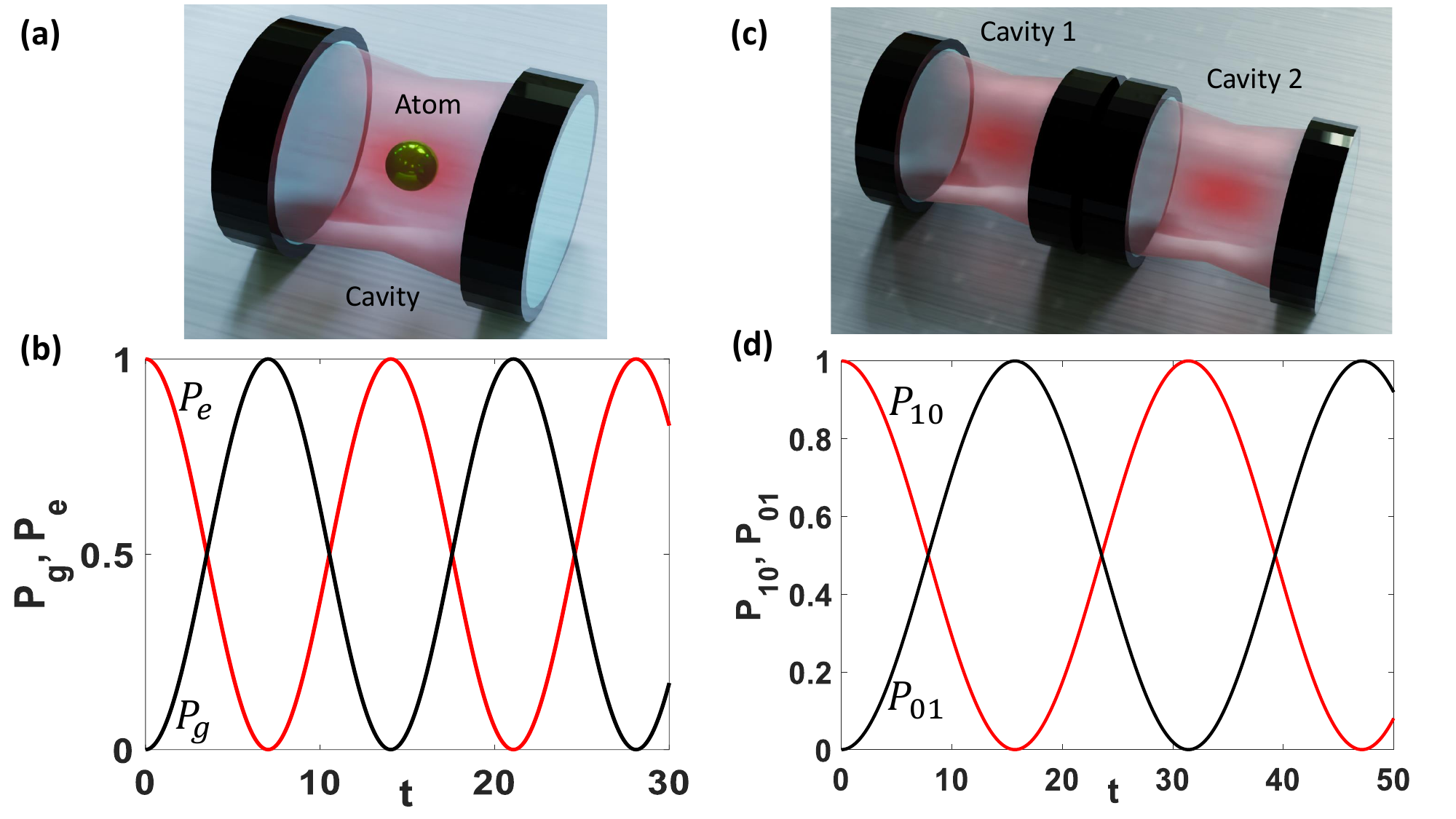}
\caption{(a) An atom is trapped inside a cavity. (b) Time evolution of the probabilities $P_{e}$ and $P_{g}$ for atom-field coupling strength $g=0.1$ and initial photon number $n=4$. (c) Two cavities Cavity 1 and Cavity 2 are placed near to each other (coupled) to exchange energy. (d) Time evolution of the probabilities $P_{10}$ and $P_{01}$ for cavity-cavity coupling strength $J=0.1$. The photon is periodically exchanged between the cavities. }  \label{AtomCavityCavity}
\end{figure}

The strength of interaction (energy exchange) between an atom with an electromagnetic field in free space is small \cite{RevModPhys.87.1379, RevModPhys.85.1083}. This is enhanced by confining them in a cavity [Fig. \ref{AtomCavityCavity}(a), \cite{RevModPhys.85.1083,Meher2022EPJP}]. The Hamiltonian for an atom-cavity system is [Ch. 4 of Ref. \cite{Gerry}, Ch. 12 of Ref. \cite{Agarwal}]
\begin{align}\label{AtomCavity}
\hat H=\frac{\hbar\omega_0}{2}\hat \sigma_z+\hbar\omega \hat a^\dagger \hat a+\hbar g(\hat \sigma_-\hat a^\dagger+\hat \sigma_+ \hat a).
\end{align}
This is the Jaynes-Cummings (JC) Hamiltonian \cite{jaynes1963comparison}.
The first and second terms are the energy operators for a two-level atom and a cavity field, respectively. We have omitted the constant $\hbar\omega/2$ from the energy operator of the cavity field since it doesn't significantly affect the time dynamics. The last term represents the interaction between field and atom, facilitating energy exchange in the form of photons. The atom absorbs a photon from the field and subsequently re-emits it into the field.\\

\fbox{
  \begin{minipage}{0.95\linewidth}  
    \textbf{Important Note:} When dealing with two physical systems in quantum mechanics, we need to be careful while writing their combined operators and states. An operator for a composite system is written as a tensor product of the operators of the individual sub-systems, that is,
    \begin{align}
      \hat C = \hat A \otimes \hat B,
    \end{align}
    where $\hat A$ and $\hat B$ are the operators corresponding to the sub-systems A and B, and $\hat C$ is an operator for the composite system.

    Similarly, a state of the combined system is also a tensor product of the states of the individual sub-systems, that is,
    \begin{align}
      \ket{\psi} = \ket{\psi_A} \otimes \ket{\psi_B}.
    \end{align}

It is important to note that we follow the order of the operators and the states to be the same, that is, we place the operator and state of the sub-system A before those of the sub-system B. This ensures that the operators of the subsystem A act meaningfully on the state of A, and similarly for the subsystem B.
  \end{minipage}
}
 
We may note the last term of Eq. \eqref{AtomCavity}, particularly the terms $\hat \sigma_-\hat a^\dagger$ and $\hat \sigma_+ \hat a$, the atomic operators are placed first and then the field operators. To follow the same order, we re-write the JC Hamiltonian as
\begin{align}\label{AtomCavityRewrite}
\hat H=\frac{\hbar\omega_0}{2} (\hat\sigma_z\otimes \hat I_f)+\hbar\omega (\hat I_a\otimes\hat a^\dagger \hat a)+\hbar g(\hat \sigma_-\otimes\hat a^\dagger+\hat \sigma_+\otimes \hat a).
\end{align} 
Here, $\hat I_a$ and $\hat I_f$  are the identity operators correspond to two-level atom and the cavity field. The dimension of $\hat I_a$ is 2 and the dimension of $\hat I_f$ depends on the dimension of the field state.  

To study the time dynamics of the system, we define the unitary operator to be
\begin{align}
\hat U=e^{-i\hat Ht/\hbar},
\end{align}
where $\hat H$ is given in Eq. \eqref{AtomCavity}. 

Let the combined initial state of atom and field be
\begin{align}
\ket{\psi_{in}}=\ket{e}\otimes\ket{n},
\end{align} 
where the atom is in its excited state $\ket{e}$ and the cavity field has precisely $n$ photons. Note that we have followed the same order for the state too (atomic state and then field state), which is necessary to have a meaningful action of the operators on the state.

For resonant case, that is, $\omega=\omega_0$, the combined state evolves under the JC Hamiltonian as [Ch. 4 of Ref. \cite{Gerry}]
\begin{align}
\ket{\psi(t)}=e^{-iHt/\hbar} \ket{\psi_{in}}=\cos(gt\sqrt{n+1})\ket{e,n}-i\sin(gt\sqrt{n+1})\ket{g,n+1},
\end{align}
where $\ket{e,n}=\ket{e}\otimes\ket{n}$ and $\ket{g,n+1}=\ket{g}\otimes\ket{n+1}$.  At $t=0$, the atom is in excited state and the cavity field is in $n$-photon state. At time $t=\pi/(2g\sqrt{n+1})$, the atom goes to its ground state by emitting a photon into the field. Then, at time $t=\pi/(g\sqrt{n+1})$, the atom comes back to its initial state $\ket{e}$ by absorbing a photon from the field. Thus, the atom will be periodically exchanging energy with the field [see Fig. \ref{AtomCavityCavity}(b)]. \\

The probability of finding the atom in excited state and the field in $n$-photon state is 
\begin{align}
P_{e}(t)=|\bra{e,n}\psi(t)\rangle|^2=\cos^2(gt\sqrt{n+1}),
\end{align}
and 
the probability of finding the atom in ground state and the field in $(n+1)$-photon state is
\begin{align}
P_{g}(t)=|\bra{g,n+1}\psi(t)\rangle|^2=\sin^2(gt\sqrt{n+1}).
\end{align}

\noindent
\textbf{Code for atom-field interaction:}
\begin{verbatim}
clear;  % Clear memory
clc;    % Clear the command window/screen
d=10;                            %dimension of the cavity field
hbar=1;
W0=1;                            %atomic frequency
Wf=1;                            %cavity field frequency
g=0.1;                           %coupling constant
A = diag(sqrt(1:d-1), 1);  % Annihilation operator
Ad=A';                      %Creation operator 
Sz=[1,0;0,-1];                     %Sigma z
Splus=[0,1;0,0];                  %sigma plus
Sminus=[0,0;1,0];                 %sigma minus
gs=[0;1];                         %ground state
es=[1;0];                         %excited state
I_a=eye(2);                       %Identity operator for the atom
I_f=eye(d);                       %Identity operator for the field
Hatom=(1/2)*hbar*W0*kron(Sz,I_f); %Atomic Hamiltonian
Hfield=hbar*Wf*kron(I_a,Ad*A);    %Field Hamiltonian
Hint=hbar*g*(kron(Splus,A)+kron(Sminus,Ad));  %interaction Hamiltonian
H=Hatom+Hfield+Hint;             %JC Hamiltonian
n=4;                               %initial number of photons in the cavity
en=kron(es,I_f(:,n+1));             % Atom in excited state, field has n photons
gn=kron(gs, I_f(:,n+2));            % Atom in ground state and field has n+1 photons
Psi=en;                           % Initial state
dt=0.1;                            %time step      
U=expm(-i*H*dt/hbar);                  % Unitary operator
 T=0:dt:30;                        % Total evolution time
for t=1:length(T);
     Pe(t)= norm(en'*Psi)^2;      %Probability of |e,n>
     Pg(t)=norm(gn'*Psi)^2;       %Probability of |g,n+1>
     Psi=U*Psi;                   %Time evolved state
     Psi=Psi/norm(Psi);           %Normalizing the state
 end
plot(T,Pe,'r',T,Pg,'k')           %plotting Pe in red colour and Pg in black colour
\end{verbatim}
\vspace{1cm}
\noindent
The output plot is shown in Fig. \ref{AtomCavityCavity}(b).
\noindent
At $t=0$, $P_e=1$ and $P_g=0$.\\
At $t=\pi/(2g\sqrt{n+1})=\pi/(2\times 0.1\times\sqrt{4+1})\approx 7$, $P_g=1$ and $P_e=0$.

Now, consider the initial state of the field to be in a coherent state and the atom in its excited state. We calculate the expectation value of the atomic operator $\sigma_z$ to be [Ch. 4 of Ref. \cite{Gerry}]
\begin{align}
W=\langle \hat\sigma_z (t)\rangle=e^{-|\alpha|^2}\sum_{n=0}^\infty \frac{|\alpha|^{2n}}{n!}\cos(2gt\sqrt{n+1}).
\end{align} 

\begin{figure}
\includegraphics[scale=0.5]{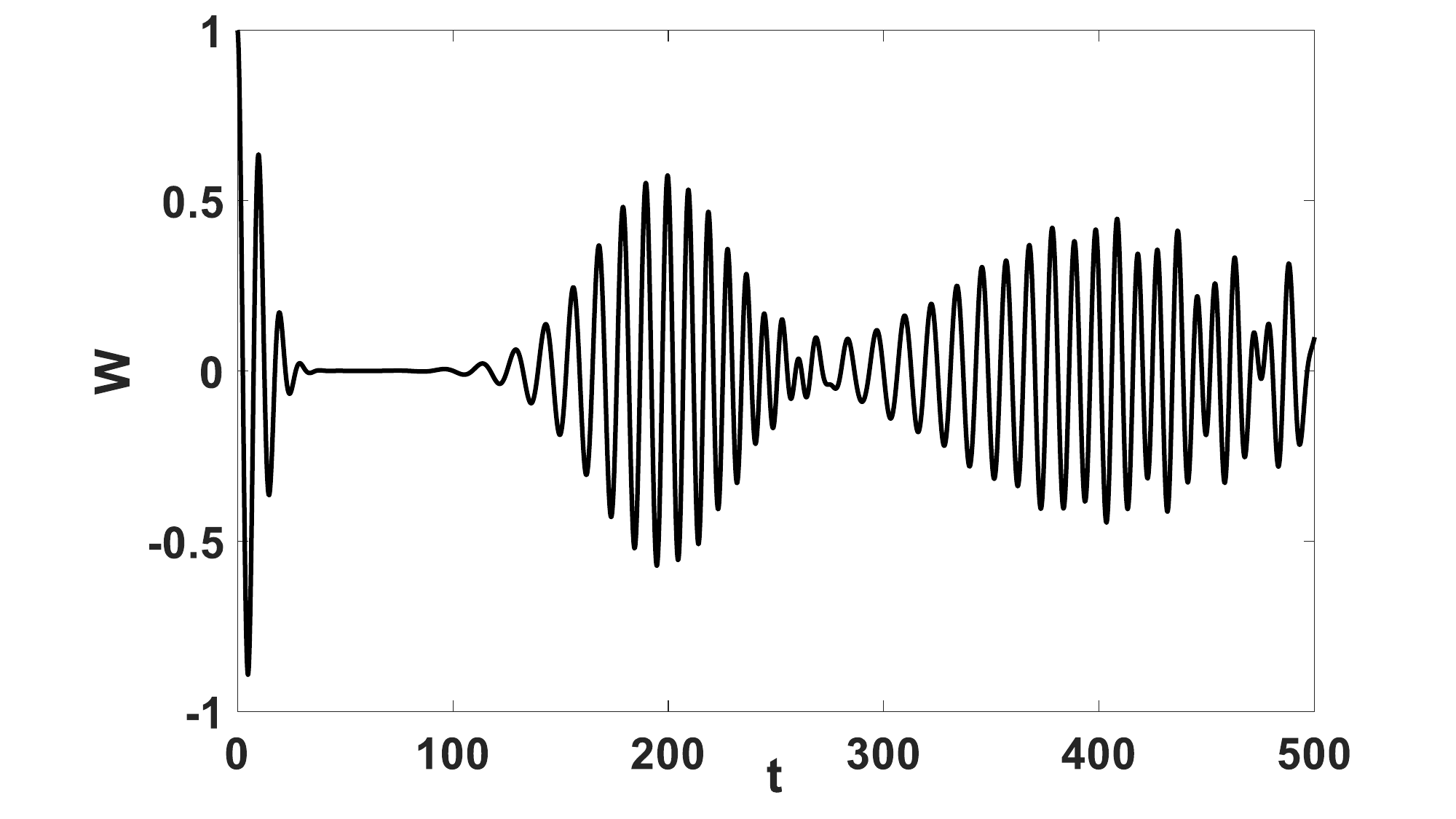}
\caption{Time evolution of $\langle \sigma_z (t)\rangle$ (atomic inversion) when the atom is in excited state and field is in a coherent state. We set $\alpha=3$ and coupling strength $g=0.1$.  }\label{PopulationInv}
\end{figure}

\noindent
\textbf{Code for calculating expectation value of atomic operator $\hat\sigma_z$ when the field is in coherent state:}
\begin{verbatim}
clear;  % Clear memory
clc;    % Clear the command window/screen
d=50;                            % dimension of the cavity field
hbar=1;
W0=1;                            %atomic frequency
Wf=1;                            %cavity field frequency
g=0.1;                           %coupling constant
A = diag(sqrt(1:d-1), 1);  % Annihilation operator
Ad=A';                      %Creation operator 
Sz=[1,0;0,-1];                     %Sigma z
Splus=[0,1;0,0];                  %sigma plus
Sminus=[0,0;1,0];                 %sigma minus
gs=[0;1];                         %ground state
es=[1;0];                         %excited state
I_a=eye(2);                       %Identity operator for the atom
I_f=eye(d);                       %Identity operator for the field
Hatom=(1/2)*hbar*W0*kron(Sz,I_f); %Atomic Hamiltonian
Hfield=hbar*Wf*kron(I_a,Ad*A);    %Field Hamiltonian
Hint=hbar*g*(kron(Splus,A)+kron(Sminus,Ad));  %interaction Hamiltonian
H=Hatom+Hfield+Hint;             %JC Hamiltonian
alpha=3;                          %coherent state amplitude
Coh=0;
for x=0:d-1;
    Coh=Coh+exp(-norm(alpha)^2/2)*alpha^x/sqrt(prod(1:x))*I_f(:,x+1);
end
Psi=kron(es,Coh);   % Initial state: atom in |e> and field in coherent state
dt=0.1;                       %time step      
U=expm(-j*H*dt);              % Unitary operator
T=0:dt:500;                   % Total evolution time
for t=1:length(T)
     W(t)=Psi'*kron(Sz,I_f)*Psi;       %Sz average (atomic inversion)
     Psi=U*Psi;
     Psi=Psi/norm(Psi);
end
plot(T,W)
\end{verbatim}
The output plot is shown in Fig. \ref{PopulationInv}. As can be seen in the plot, the oscillation of $\langle \sigma_z (t)\rangle$ (atomic inversion) collapses (W=0) for a duration of time and then revives. For more details on this interesting observation, specifically the collapse and revival of $W$, we recommend the readers to see the standard textbooks \cite{Gerry,Agarwal, Scully}.

\section{Two-mode field}
If two electromagnetic fields differ in either frequency, propagation direction, or polarization, they are referred to as two-mode fields. In such cases, we use different field operators to represent them. Let the annihilation operators for a two-mode field be $\hat a_1$ and $\hat a_2$, and  the creation operators be $\hat a_1^\dagger$ and $\hat a_2^\dagger$.
\subsection{Coupled cavities (energy exchange between two field modes)}
These two fields exchange energy if we confine them in two coupled cavities and placed near to each other [Fig. \ref{AtomCavityCavity}(c)]. If the two cavities have same resonance frequencies, then the Hamiltonian for this configuration is \cite{Meher2022EPJP}
\begin{align}
\hat H=\hbar\omega \hat a_1^\dagger \hat a_1 +\hbar\omega \hat a_2^\dagger \hat a_2+\hbar J(\hat a_1^\dagger \hat a_2+\hat a_1 \hat a_2^\dagger),
\end{align}
where $\omega$ is the resonant frequencies of the both the cavities, and $J$ is the coupling strength between them. 

We consider a simple case: One photon is in the first cavity and the second cavity is in vacuum (no photons). Then, the initial state is
\begin{align}
\ket{\psi_{in}}=\ket{1,0}.
\end{align}
The state at a later time $t$ is
\begin{align}
\ket{\psi(t)}=e^{-i\hat Ht/\hbar}\ket{\psi_{in}}=\cos Jt\ket{1,0}-i\sin Jt\ket{0,1}.
\end{align}
The probability of finding the photon in the first cavity is 
\begin{align}
P_{10}(t)=|\bra{1,0}\psi(t)\rangle |^2=\cos^2 Jt,
\end{align}
and the probability of finding the photon in the second cavity is
\begin{align}
P_{01}(t)=|\bra{0,1}\psi(t)\rangle |^2=\sin^2 Jt.
\end{align}

\noindent
\textbf{Code for coupled-cavity dynamics:}

\begin{verbatim}
clear;  % Clear memory
clc;    % Clear the command window/screen
d=10;                            %dimension of the cavity field
hbar=1;
W1=1;                            %Resonance frequency of the first cavity
W2=1;                            %Resonance frequency of the first cavity
J=0.1;                           %inter-cavity coupling constant
A = diag(sqrt(1:d-1), 1);  % Annihilation operator
Ad=A';                      %Creation operator 
I_f=eye(d);                 %Identity operator for the fields
H=hbar*W1*kron(Ad*A,I_f)+hbar*W2*kron(I_f,Ad*A)+hbar*J*(kron(Ad,A)+kron(A,Ad));
S10=kron(I_f(:,2),I_f(:,1));  %First cavity in |1> photon and second cavity in |0>
S01=kron(I_f(:,1),I_f(:,2));  %Second cavity in |1> and first cavity in |0>
Psi=S10;                         %Initial state
dt=0.1;                          %time step      
U=expm(-i*H*dt/hbar);            % Unitary operator
T=0:dt:50;                       % Total evolution time
for t=1:length(T)
   P10(t)=norm(S10'*Psi)^2;    %Prob. of finding the photon in first cavity
   P01(t)=norm(S01'*Psi)^2;    %Prob. of finding the photon in second cavity
   Psi=U*Psi;                  %Time evolution
   Psi=Psi/norm(Psi);          %Normalization
end
plot(T,P10,'r',T,P01,'k')         %Plotting in red and black
\end{verbatim}
The output plot is shown in Fig. \ref{AtomCavityCavity}(d). The photon, initially which was in first cavity, is completely transferred to the second cavity at time $t=\pi/2J$.\\  

One may consider the general case: The first cavity contains $N-n$ photons and the second cavity contains $n$ photons, such that, the total number of photons is $N$, then the time evolved state is \cite{Meher2017ScRep}
\begin{align}
\ket{\psi(t)}=e^{-i\hat Ht/\hbar}\ket{N-n,n}=e^{-iN\omega t}\sum_{k,k'=0}^{N-n,n}&~^{N-n}C_k~^nC_{k'} (\cos Jt)^{N-(k+k')} (-i\sin Jt)^{k+k'}\nonumber\\
&\times\sqrt{\frac{~^NC_n}{~^NC_{n+k-k'}}}\ket{N-(n+k-k'),n+k-k'},
\end{align}
where $~^pC_q=\tfrac{p!}{(p-q)!q!}$ is the binomial coefficient. 


\subsection{Beam splitter transformation}
\begin{figure}
\includegraphics[scale=0.5]{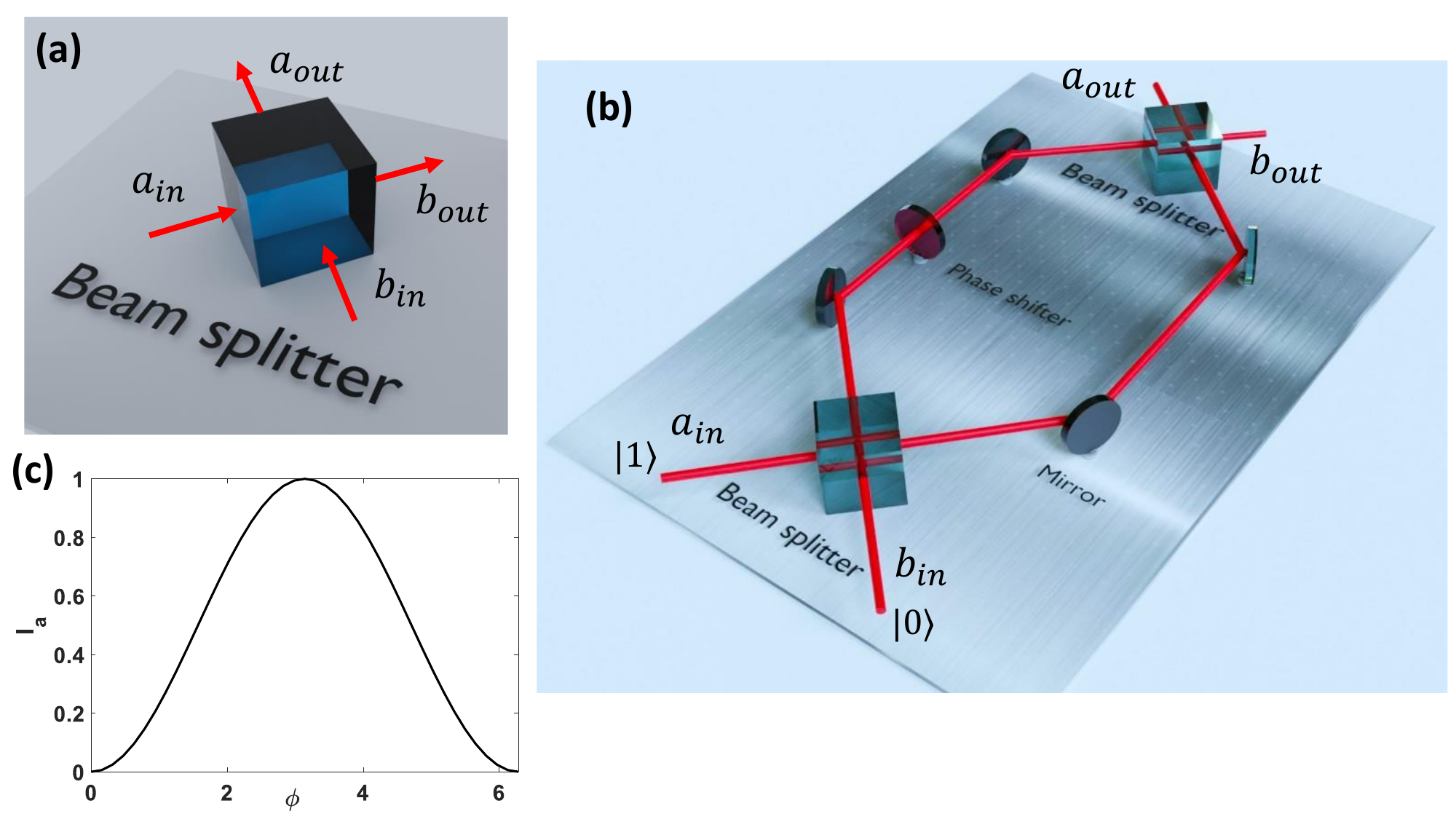}
\caption{(a) A beam splitter: two input ports $a_{in}, b_{in}$ and two output ports $a_{out},b_{out}$. (b) A schematic of Mach-Zehnder interferometer consisting of two beam splitters, four mirrors and a phase-shifter. (c) Interference pattern: output intensity at $a_{out}$ as a function of $\phi$.}\label{MZIBS}
\end{figure}

A beam splitter has two input ports and two output ports [see Fig. \ref{MZIBS}(a)]. When a beam of light is directed into one of the input ports, the beam splitter divides it into two separate beams. The intensity of the resulting output beams is determined by the splitting ratio of the beam splitter. In case of a 50:50 ratio beam splitter, the intensities of the output beams are equal.

If we consider a quantum field (nonclassical light), suppose a number state, input to one of the ports of the beam splitter, then the beam splitter will distribute the photons in two outputs with different probabilities. For instance, if the input is number state $\ket{5}$ and the other port is in vacuum (no photon), the possible outputs are: $\ket{5,0},\ket{4,1},\ket{3,2},\ket{2,3},\ket{1,4},\ket{0,5}$.  The probabilities of these states depend on the beam splitter ratio. However, once we do a measurement (photon detection) at the output, we realize one of them.

Consider a beam splitter as shown in Fig. \ref{MZIBS}(a) with its unitary transformation [Ch. 6 of Ref. \cite{Gerry}, Ch. 5 of Ref. \cite{Agarwal}]
\begin{align}
\hat U_{BS}=e^{i\theta(\hat a^\dagger \hat b+\hat a \hat b^\dagger)},
\end{align} 
where $\hat a$ and $\hat b$ are the annihilation operators for mode-a and mode-b, respectively. The parameter $\theta$ decides the beam splitting ratio, that is, $\cos^2\theta$ and $\sin^2\theta$ are the reflectivity and transmissivity of the beam splitter, respectively.  We consider the beam splitter to be 50:50 by taking $\theta=\pi/4$. As a result, it splits the beam into two with equal intensities.

Let the input be $\ket{n,0}$, where $n$ photons are in mode-a and the mode-b is in vacuum. Thus, the input intensity is $\langle \hat a^\dagger \hat a\rangle=n$. The output state will be a superposition of all possible combinations generated from the distribution of $n$ photons in two modes [Ch. 5 of Ref. \cite{Agarwal}]:
\begin{align}\label{BStransform}
\ket{\psi_{out}}=\hat U_{BS}\ket{n,0}=\frac{1}{\sqrt{2^n}}\sum_{k=0}^n i^k\sqrt{\frac{n!}{(n-k)!k!}}\ket{n-k,k}.
\end{align}
The average number of photons ($\sim$ intensity) at the output port $a_{out}$ is
\begin{align}
I_a=\bra{\psi_{out}}(\hat a^\dagger \hat a\otimes \hat I_f)\ket{\psi_{out}}=\frac{n}{2},
\end{align}
and at the output port $b_{out}$ is
\begin{align}
I_a=\bra{\psi_{out}}(\hat I_f\otimes \hat b^\dagger \hat b)\ket{\psi_{out}}=\frac{n}{2}.
\end{align}
After the 50:50 beam splitter transformation, each output gets half of its input intensity.

For $n=2$, the input state is $\ket{2,0}$. The output state can be calculated using Eq. \eqref{BStransform} to be
\begin{align}
\ket{\psi_{out}}=\hat U_{BS}\ket{2,0}=\frac{1}{2}\ket{2,0}+\frac{i}{\sqrt{2}}\ket{1,1}-\frac{1}{2}\ket{0,2}.
\end{align}
Therefore, the probabilities of getting $\ket{2,0}, \ket{1,1}$ and $\ket{0,2}$ are 1/4, 1/2 and 1/4, respectively. The average number of photons ($\sim$ intensity) at the output port $a_{out}$ is
\begin{align}
I_a=\bra{\psi_{out}}(\hat a^\dagger \hat a\otimes I_f)\ket{\psi_{out}}=1,
\end{align}
and at the output port $b_{out}$ is
\begin{align}
I_b=\bra{\psi_{out}}(I_f\otimes \hat b^\dagger \hat b)\ket{\psi_{out}}=1.
\end{align}

\fbox{
  \begin{minipage}{0.95\linewidth}  
  For the input $\ket{n,m}$, the output state after the beam splitter is [Ch. 5 of Ref. \cite{Agarwal}]
\begin{align}
\hat U_{BS}\ket{n,m}=\sum_{k,k'=0}^{n,m} ~^nC_k~^mC_{k'} & (\cos\theta)^{m+k-k'} (i\sin\theta)^{n-k+k'}\nonumber\\
&\times\sqrt{\frac{(k+k')!(n+m-k-k')!}{n! m!}}\ket{k+k',n+m-k-k'}.
\end{align}
For 50:50 beam splitter, use $\theta=\pi/4$.   
  \end{minipage}
}

\vspace{1cm}
\noindent
\textbf{Code for beam splitter transformation:}
\begin{verbatim}
clear;  % Clear memory
clc;    % Clear the command window/screen
d=10;                            %dimension of the cavity field
A = diag(sqrt(1:d-1), 1);  % Annihilation operator
Ad=A';                      %Creation operator 
AdA=Ad*A;                   %Number operator         
theta=pi/4;                %beam splitter parameter
UBS=expm(i*theta*(kron(Ad,A)+kron(A,Ad)));
I_f=eye(d);                 %Identity operator for the fields
S20=kron(I_f(:,3),I_f(:,1));  %mode-a has two photons and mode-b is in vacuum
S11=kron(I_f(:,2),I_f(:,2));  %mode-a has one photon and mode-b has one photon
S02=kron(I_f(:,1),I_f(:,3));  %mode-b has two photons and mode-a is in vacuum
Psi=S20;                    %Input state
PsiOut=UBS*Psi;               %output state         
P20=norm(S20'*PsiOut)^2      %Probability of 20
P11=norm(S11'*PsiOut)^2      %Probability of 11
P02=norm(S02'*PsiOut)^2      %Probability of 02
AdAout=PsiOut'*kron(AdA,I_f)*PsiOut    %Intensity at port-a
BdBout=PsiOut'*kron(I_f,AdA)*PsiOut    %Intensity at port-b
\end{verbatim}
\vspace{1cm}
\noindent
The outputs are:
\begin{verbatim}
P20 =
    0.2500
P11 =
    0.5000
P02 =
    0.2500    
AdAout =
     1
BdBout =
     1
\end{verbatim}
\subsection{Mach-Zehnder interferometer}
The Mach-Zehnder interferometer (MZI) is one of the simplest devices that demonstrates interference of an electromagnetic field at a single photon level [Ch. 6 of \cite{Gerry}]. It consists of two beam splitters, a few mirrors, and a phase shifter [See Fig. \ref{MZIBS}(b)]. The first beam splitter divides the input field into two. A phase shifter in one of the arms/paths introduces a phase shift to the field passing through it, and the final beam splitter combines both the beams. Each mirror contributes a phase shift of $\pi/2$ in both the paths, amounting to an irrelevant global phase, that we omit.

To observe interference pattern at the output ports of the MZI [Fig. \ref{MZIBS}(b)], we consider a single-photon state input to the port $a_{in}$ and a vacuum state to the other port $b_{in}$. Therefore, the input state is
\begin{align}
\ket{\psi_{in}}=\ket{1,0}.
\end{align}
The state after the first beam splitter is (from Eq. \eqref{BStransform})
\begin{align}
\ket{\psi_1}=\hat U_{BS}\ket{1,0}=\frac{1}{\sqrt{2}}(\ket{1,0}+i\ket{0,1}),
\end{align}
The phase shifter, placed in the left arm, introduces a phase $\phi$ if a photon passes through it. In general, it shifts a phase of $n\phi$ if $n$ photons pass through it. Thus, the unitary operator for the phase-shifter is $e^{i\phi \hat a^\dagger \hat a}$.  In our case, $n=1$ and therefore, we have
\begin{align}
\ket{\psi_2}=\frac{1}{\sqrt{2}}(e^{i\phi}\ket{1,0}+i\ket{0,1}).
\end{align} 
The final beam splitter transforms the above state to
\begin{align}
\ket{\psi_{out}}&=\frac{1}{\sqrt{2}}\left(e^{i\phi}\frac{1}{\sqrt{2}}(\ket{1,0}+i\ket{0,1})+i\frac{1}{\sqrt{2}}(i\ket{1,0}+\ket{0,1})\right) \nonumber\\
&=\frac{1}{\sqrt{2}}\left((e^{i\phi}-1)\ket{1,0}+i (e^{i\phi}+1)\ket{0,1} \right),
\end{align}
where we applied the beam splitter transformation on each components $\ket{1,0}$ and $\ket{0,1}$.

The probability of finding the photon at $a_{out}$ is 
\begin{align}
P_{10}=|\bra{1,0}\psi_{out}\rangle|^2=\frac{1}{2}(1-\cos\phi),
\end{align}
and the probability of finding the photon at $b_{out}$ is 
\begin{align}
P_{01}=|\bra{0,1}\psi_{out}\rangle|^2=\frac{1}{2}(1+\cos\phi).
\end{align}
The average number of photons ($\sim$ intensity) at the output port $a_{out}$ is
\begin{align}
I_a=\bra{\psi_{out}}(a^\dagger a\otimes I_f)\ket{\psi_{out}}=\frac{1}{2}(1-\cos\phi),
\end{align}
and at the output port $b_{out}$ is
\begin{align}
I_b=\bra{\psi_{out}}(I_f\otimes b^\dagger b)\ket{\psi_{out}}=\frac{1}{2}(1+\cos\phi).
\end{align}
It is to be noted that, as the input contains one photon, the probability and average number of photons are equal. Importantly, all the above calculated quantities oscillate between 1 and 0 when $\phi$ varies. Thus, the phase variation inside the interferometer displays an interference pattern at the output (see Fig. \ref{MZIBS}(c))  \\

\noindent
\textbf{Code for the Mach-Zehnder interferometer:}
\begin{verbatim}
clear;  % Clear memory
clc;    % Clear the command window/screen
d=10;                            %dimension of the cavity field
I_f=eye(d);                 %Identity operator for the fields 
A = diag(sqrt(1:d-1), 1);  % Annihilation operator
Ad=A';                      %Creation operator 
AdA=Ad*A;                   %Number operator         
theta=pi/4;                %beam splitter parameter
UBS=expm(i*theta*(kron(Ad,A)+kron(A,Ad)));    
S10=kron(I_f(:,2),I_f(:,1));  %mode a has one photon and mode b is in vacuum
S01=kron(I_f(:,1),I_f(:,2));  %mode b has one photon and mode a is in vacuum
Psi=S10;                     %Input state |10>
phir=0:pi/20:2*pi;   %phase shift running from 0 to 2*pi
for x=1:length(phir)
    phi=phir(x);       %phase shift
    Uphi=expm(i*phi*kron(AdA,I_f));  %phase shift unitary operator
    PsiOut=UBS*Uphi*UBS*Psi;        %Output state after MZI       
    P10(x)=norm(S10'*PsiOut)^2;      %Probability of 10	
	P01(x)=norm(S01'*PsiOut)^2;     %Probability of 01
    AdAout(x)=PsiOut'*kron(AdA,I_f)*PsiOut;    %Intensity at output port-a
    BdBout(x)=PsiOut'*kron(I_f,AdA)*PsiOut;    %Intensity at output port-b
end
plot(phir,AdAout,'k')
\end{verbatim}
The plot for the output intensity at $a_{out}$ is shown in Fig. \ref{MZIBS}(c).

\section{Dissipative atom-field dynamics}
Complete isolation of a system from its surrounding is not possible. Interaction between
a system and the environment leads to a dissipative dynamics of the system.  There are several approaches for incorporating dissipation in quantum systems. Among them, the Lindblad master equation and Monte-Carlo wavefunction method are the most used approaches in quantum optics that we discuss here. 
\subsection{Lindblad master equation}
In this method, the evolution equation for the reduced density matrix of a system is obtained by tracing over the states of the reservoir (zero temperature). For an atom-cavity system, the master equation is \cite{Carmichael}
\begin{align}\label{master}
\frac{d}{d t}\rho_s=\frac{1}{i\hbar}[\hat H,\rho]+\mathcal{L}_{cavity}(\rho)+\mathcal{L}_{atom}(\rho),
\end{align}
where 
\begin{align}
\mathcal{L}_{cavity}(\rho)&=\frac{\kappa}{2}(2\hat a\rho \hat a^\dagger-\rho \hat a^\dagger \hat a-\hat a^\dagger\hat a\rho),\\
\mathcal{L}_{atom}(\rho)&=\frac{\gamma}{2}(2\hat \sigma_-\rho \hat \sigma_+-\rho \hat \sigma_+ \hat \sigma_--\hat \sigma_+\hat\sigma_-\rho),
\end{align}
are the Lindblad superoperators \cite{lindblad1976generators}. Here, $\gamma$ and $\kappa$ are the atomic and cavity decay rates, respectively. The Hamiltonian $\hat{H}$ is given in Eq. \eqref{AtomCavityRewrite}.

The time evolution of the expectation value of any system operator can be calculated as 
\begin{align}
\langle \hat O(t)\rangle=\text{Tr}(\hat O\rho_s(t)).
\end{align}

\begin{figure}
\includegraphics[scale=0.5]{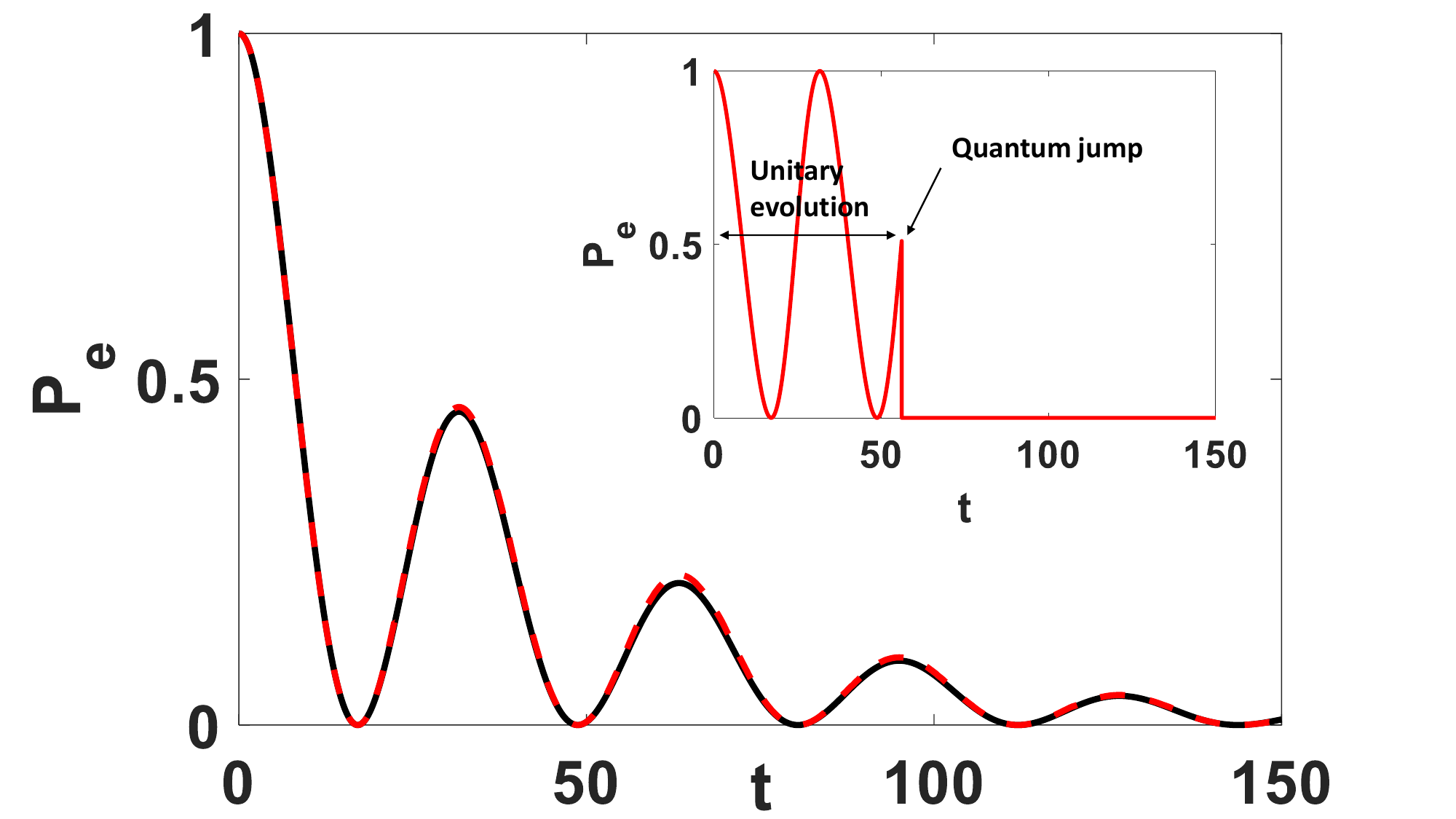}
\caption{The probability of finding the atom in its excited state, calculated using master equation approach (black continuous line), as a function of time $t$ in the presence of cavity dissipation. The result is compared with the Monte-Carlo wavefunction method (red dashed line). We set $\kappa=0.05, \gamma=0,$ and $g=0.1$. For MCWF, the number of realizations is $N=5000$. The inset shows a single trajectory generated using MCWF method. A quantum jump occurs at $t\sim 56$. }\label{masterMCWF}
\end{figure}

In the following code, for simplicity, we consider only the cavity is dissipative. So, we take non-zero cavity decay rate $\kappa$, and the atomic decay rate to be $\gamma=0$. To solve the evolution equation given in Eq. \eqref{master}, we use Runge-Kutta method (RK4) by considering the initial state to be $\ket{e,0}$. We calculate the probability of finding the atom in excited state at various instant of time through
\begin{align}
P_e(t)=\bra{e}\rho_s(t)\ket{e}.
\end{align}
and shown in Fig. \ref{masterMCWF} (black continuous line). The energy of the atom is dissipated due to cavity decay.

\vspace{1cm}
\noindent
\textbf{Code for master equation:}
\begin{verbatim}
clear;  % Clear memory
clc;    % Clear the command window/screen
d=10;                            % dimension of the cavity field
hbar=1;
W0=1;                            %atomic frequency
Wf=1;                            %cavity field frequency
g=0.1;                           %coupling constant
kappa=0.05;                     %cavity decay rate
A = diag(sqrt(1:d-1), 1);  % Annihilation operator
Ad=A';                      %Creation operator 
Sz=[1,0;0,-1];                     %Sigma z
Splus=[0,1;0,0];                  %sigma plus
Sminus=[0,0;1,0];                 %sigma minus
gs=[0;1];                         %ground state
es=[1;0];                         %excited state
I_a=eye(2);                       %Identity operator for the atom
I_f=eye(d);                       %Identity operator for the field
Hatom=(1/2)*hbar*W0*kron(Sz,I_f); %Atomic Hamiltonian
Hfield=hbar*Wf*kron(I_a,Ad*A);    %Field Hamiltonian
Hint=hbar*g*(kron(Splus,A)+kron(Sminus,Ad));  %interaction Hamiltonian
H=Hatom+Hfield+Hint;             %JC Hamiltonian
Psi=kron(es,I_f(:,1));           %Atom in excited state and cavity in vacuum
Rho=Psi*Psi';                    %initial density matrix
dt=0.1;                         %time step      
T=0:dt:150;                     % Total evolution time

%Re-define the annihilation and creation operators
%to match the total dimension (dim(atom)*dim(field))
An=kron(I_a,A);
Adn=kron(I_a,Ad);

% The master equation is solved by Runge-Kutta method (RK4)
for t=1:length(T)
    Pe(t)= kron(es,I_f(:,1))'*Rho* kron(es,I_f(:,1));  %Probability of atom in |e>    
    K1=-i*(H*Rho-Rho*H)+kappa/2*(2*An*Rho*Adn-Adn*An*Rho-Rho*Adn*An);
    Rho1=Rho+1/2*dt*K1;
    K2=-i*(H*Rho1-Rho1*H)+kappa/2*(2*An*Rho1*Adn-Adn*An*Rho1-Rho1*Adn*An);
    Rho2=Rho+1/2*dt*K2;
    K3=-i*(H*Rho2-Rho2*H)+kappa/2*(2*An*Rho2*Adn-Adn*An*Rho2-Rho2*Adn*An);
    Rho3=Rho+dt*K3;
    K4=-i*(H*Rho3-Rho3*H)+kappa/2*(2*An*Rho3*Adn-Adn*An*Rho3-Rho3*Adn*An);    
    Rho=Rho+1/6*(K1+2*K2+2*K3+K4)*dt;
end
plot(T,Pe)
\end{verbatim}

\subsection{Monte-Carlo wavefunction method}
When the system has finite number
of levels, Monte-Carlo wavefunction (MCWF) approach can be used for investigating the effect of dissipation \cite{molmer1993monte}. This method is also known as \textquoteleft quantum jump
approach'. To study the effect of dissipation using this method, the
quantum state is evolved under a non-Hermitian Hamiltonian and quantum jumps
are randomly introduced. This evolution of the quantum state forms a quantum trajectory, and
ensemble average over many realizations of such trajectories reproduces the
results of the master equation.\\

\noindent
Steps for MCWF approach in atom-cavity system (for a general case, see Ref. \cite{molmer1993monte}):
\begin{itemize}
\item Consider initially the atom-cavity system be in a normalized state $\ket{\psi(0)}$.  In order to decide the state at later time $\delta t$, the state $\ket{\psi(0)}$ evolves under a non-Hermitian Hamiltonian
\begin{align}
\hat H_{NH}=\hat H -\frac{i\hbar\kappa}{2}\hat a^\dagger \hat a,
\end{align}
where $\hat{H}$ is given in Eq. \eqref{AtomCavityRewrite}. Here, we consider only the cavity is dissipating. For $\delta t\ll 1$, the state at a time $\delta t$ is
\begin{align}
\ket{\psi(\delta t)}=\left(\hat I-\frac{i\delta t}{\hbar}\hat{H}_{NH}\right)\ket{\psi(0)},
\end{align}
where $\hat I=\hat I_a\otimes I_f$ is the identity matrix for the atom-cavity system. As the Hamiltonian $\hat H_{NH}$ is non-Hermitian, the evolution does not preserve the norm and the norm will be less than 1. So, the missing norm is
\begin{align}
\delta p=1-\langle \psi(\delta t) \ket{\psi(\delta t)},
\end{align}
which should be much less than 1. 

\item Once $\delta p$ is calculated, it will be compared with a random number $r$ chosen from a uniform distribution between 0 and 1. If $r > \delta p$, no jump occurs and the state $\ket{\psi(\delta t)}$ will be normalized. And if $r < \delta p$, a quantum jump occurs, and the normalized state at $\delta t$ will be
\begin{align}
\ket{\psi(\delta t)} =\frac{\hat a\ket{\psi(0)}}{\sqrt{\langle \psi(0)|\hat a^\dagger \hat a |\psi(0)\rangle}}.
\end{align}

\item We then calculate expectation value of any observable at time $\delta t$ through
\begin{align}
\langle \hat O (\delta t) \rangle=\langle \psi(\delta t)|\hat O \ket{\psi(\delta t)}.
\end{align}

\item To get the state at $2\delta t$, we need to follow the same procedure above: evolving under a non-Hermitian Hamiltonian, calculate the missing norm and compare with a random number to decide the state at a later time. This procedure has to be followed up to a time $T=n\delta t$. This forms a single trajectory. But, one has to get many such quantum trajectories from the initial state $\ket{\psi(0)}$, and ensemble average over many such trajectories produces
the exact evolution for the initial quantum state.  
\end{itemize}

The following code considers the dissipation of an atom through cavity $(\kappa\neq 0, \gamma=0$). The probability of finding the atom in excited state after each MCWF step is calculated through
\begin{align}
P_e(\delta t)=|\bra{e}\psi(\delta t)\rangle|^2,
\end{align}
and taken averages over many realizations. The result is compared with the results of the master equation approach in Fig. \ref{masterMCWF}. We take 5000 realizations. The inset shows one of the MCWF trajectories in which a quantum jump (photon loss) occurs at $t\sim 56$.  After the quantum jump, the atom and cavity reach to their ground state. If the results from MCWF method do not match with that of the master equation approach, one needs to increase the number of realizations and decrease the time step $\delta t$. However, for such cases, the MCWF method becomes computationally much more expensive than the Lindblad master equation method. \\

\vspace{1cm}
\noindent
\textbf{Code for MCWF:}
\begin{verbatim}
clear;  % Clear memory
clc;    % Clear the command window/screen
d=10;                            % dimension of the cavity field
hbar=1;
W0=1;                            %atomic frequency
Wf=1;                            %cavity field frequency
g=0.1;                           %atom-cavity coupling constant
kappa=0.05;                     %cavity decay rate
A = diag(sqrt(1:d-1), 1);  % Annihilation operator
Ad=A';                      %Creation operator 
AdA=Ad*A;                   %number operator
Sz=[1,0;0,-1];                     %Sigma z
Splus=[0,1;0,0];                  %sigma plus
Sminus=[0,0;1,0];                 %sigma minus
gs=[0;1];                         %ground state
es=[1;0];                         %excited state
I_a=eye(2);                       %Identity operator for the atom
I_f=eye(d);                       %Identity operator for the field
Hatom=(1/2)*hbar*W0*kron(Sz,I_f); %Atomic Hamiltonian
Hfield=hbar*Wf*kron(I_a,Ad*A);    %Field Hamiltonian
Hint=hbar*g*(kron(Splus,A)+kron(Sminus,Ad));  %interaction Hamiltonian
H=Hatom+Hfield+Hint;             %JC Hamiltonian
HNH=H-i*kappa/2*kron(I_a,AdA);           %non-Hermitian Hamiltonian
dt=0.001;                         %time step is taken to be very small    
T=0:dt:150;                       % Total evolution time
%Re-define the annihilation and creation operators 
%to match the total dimension (dim(atom)*dim(field))
An=kron(I_a,A);
Adn=kron(I_a,Ad);
N=100;   %number of realization
PeAvg=0;    %initialization to zero
for x=1:N 
Psi=kron(es,I_f(:,1));     %Initial state: Atom in |e> and cavity in |0>
for t=1:length(T)
    Pe(t)=norm(kron(es,I_f(:,1))'*Psi)^2;  %Probability of atom in |e> and field |0>
    PsiNH=(kron(I_a,I_f)-i*dt*HNH)*Psi;
    dp=1-PsiNH'*PsiNH;   %missing norm
     r=rand;  %random number
    if dp<r  %comparing with a random number
       Psi=PsiNH/norm(PsiNH);    %normalized the state  
    else
       PsiJump=kron(I_a,A)*PsiNH; %quantum jump occurs
       Psi=PsiJump/norm(PsiJump); %normalize
    end
end
PeAvg=PeAvg+Pe;    %adding the probability of all the realizations to find the average
x
end
plot(T,PeAvg/N,'k')  %Plot after averaging to number of realizations
\end{verbatim}

\section{Conclusion}
We have presented a beginner-level numerical guide written in MATLAB, which can serve as a basic toolkit for addressing research problems on quantum optics and related areas such as quantum many-body physics, quantum information processing and quantum computation.  The provided codes will be highly useful for graduate students and researchers embarking on their careers in these fields. The codes can be easily extended to tackle problems involving high-dimensional matrices. Importantly, they can be executed in any version of MATLAB without requiring pre-installed packages.\\

\noindent
\textbf{Acknowledgment:} The author acknowledges Prof. S. Sivakumar, Dr. Saikat Sur, Dr. Pritam Chattopadhyay, and Dr. Binay K. Sahu for useful suggestions and discussions.
\appendix
\section{Role of dimension of the field}\label{Dimension}
The following code gives a coherent state:
\begin{verbatim}
clear;  % Clear memory
clc;    % Clear the command window/screen
d=10; %dimension of the field
I=eye(d); 
alpha=2; %Amplitude of the coherent state
Coh=0;  %initialization
for x=0:d-1;
Coh=Coh+exp(-norm(alpha)^2/2)*alpha^x/sqrt(prod(1:x))*I(:,x+1);
end
Coh    %it will display the coherent state
N_c=norm(Coh)  %checking norm which must be 1
\end{verbatim} 
In the above program, we have taken the amplitude of the coherent state to be alpha=2, by keeping the dimension of the field to be d=10. Now, if we run, we get the output to be:
\begin{verbatim}
Coh =

    0.1353
    0.2707
    0.3828
    0.4420
    0.4420
    0.3953
    0.3228
    0.2440
    0.1725
    0.1150


N_c =

    0.9959   
\end{verbatim}
We see that the norm of the state is not equal to 1, also the last element of the column matrix is not close to zero. So, this will give error if we do any further calculation using this state. Thus, the dimension of the field has to be increased. Check by taking $d=15$, we will get the norm N\_c to be very close to 1.


%

\end{document}